# Phase-space representation of completely positive quantum operation: Invertible subdynamics of two-mode squeezed state

Sumiyoshi Abe[1,2] and Yasuyuki Matsuo[1]

[1]*Department of Physical Engineering, Mie University, Mie 514-8507, Japan*

[2]*Institute of Physics, Kazan Federal University, Kazan 420008, Russia*

**Abstract**    Completely-positive quantum operations are frequently discussed in the contexts of statistical mechanics and quantum information. They are customarily given by maps forming positive operator-values measures. To intuitively understand physical meanings of such abstract operations, the method of phase-space representations is examined. This method enables one to grasp the operations in terms of the classical statistical notions. As an example of physical importance, here, the phase-space representation of the completely-positive quantum operation arising from the single-mode subdynamics of the two-mode squeezed vacuum state, which maps from the vacuum state at vanishing temperature to mixed states with perfect decoherence including the thermal state, is studied. It is found in the $P$ representation that remarkably this operation is invertible, implying that coherence lost by the quantum operation can be recovered.

# I. INTRODUCTION

The concept of generalized measurements and associated nonunitary operations play important roles in statistical mechanics and quantum information. Consider one such operation, $\Lambda$, which maps from an "initial" density matrix $\rho_0$ to another $\rho$, that is, $\rho_0 \to \rho = \Lambda(\rho_0)$. It is natural in view of the principle of quantum mechanic to impose the conditions that it is linear, completely positive, and trace-preserving. Then, the most general form of such $\Lambda$ is given by the Kraus representation [1-4]

$$\rho = \Lambda(\rho_0) = \sum_n K_n \rho_0 K_n^\dagger, \qquad (1)$$

where the matrices $K_n$'s obey

$$\sum_n K_n^\dagger K_n = I \qquad (2)$$

with $I$ being the identity operator. Eq. (2) ensures that $\Lambda$ is, in fact, trace-preserving: $\operatorname{Tr}\Lambda(\rho_0) = \operatorname{Tr}\rho_0 (=1)$. The set, $\{K_n^\dagger K_n\}_n$, forms a positive operator-valued measure (commonly abbreviated as POVM).

Now, although the operation in Eq. (1) is convenient for describing nonunitary quantum subdynamics [1-4], it is apparently formal and abstract. It may be physically desirable to be able to perceive it in a more intuitive manner. In this sense, the method of phase-space representations [5] is expected to offer a useful tool, since it helps one to grasp formal quantum operations in terms of the classical statistical notions.

In this paper, we develop the phase-space approach to the completely-positive quantum operation of the form in Eq. (1). We present the integral transformation in phase space associated with Eq. (1). In particular, we consider the operation arising from the subdynamics of the two-mode squeezed vacuum state, which realizes perfect decoherence and generates the thermal state, and analyze its phase-space representations. We derive an explicit formula for the transformation kernel and find a remarkabke fact that the transformation is of the convolution type and therefore is invertible. The restorability of coherence may be relevant to, e.g., error correction in quantum information processing.

Throughout this article, both $\hbar$ and the Boltzmann constant, $k_B$, are set equal to unity for the sake of simplicity.

## II. REDUCTION OF TWO-MODE SQUEEZED VACUUM AND ASSOCIATED QUANTUM OPERATION

In this section, we discuss the completely-positive quantum operation derived from the two-mode squeezed vacuum state. Consider a two-mode radiation field. Its quadratures, $X_A$, $X_B$, $Y_A$, $Y_B$, are given in terms of the creation and annihilation operators, $a^\dagger$, $b^\dagger$ and $a$, $b$ as follows: $X_A = a^\dagger + a$, $X_B = b^\dagger + b$, $Y_A = i(a^\dagger - a)$, $Y_B = i(b^\dagger - b)$. The basic commutation relations are: $[a, a^\dagger] = I_A$, $[a, a] = [a^\dagger, a^\dagger] = 0$, $[b, b^\dagger] = I_B$, $[b, b] = [b^\dagger, b^\dagger] = 0$, $[a, b^\dagger] = 0$ and so on, where $I_A$ and $I_B$ are the identity operators in the spaces of the modes A and B, respectively. The two-mode squeezed vacuum is defined as follows:

$$|\theta\rangle = U(\theta)|0\rangle_A |0\rangle_B, \tag{3}$$

$$U(\theta) = \exp\left[\theta\left(a^\dagger b^\dagger - ab\right)\right]. \tag{4}$$

$|0\rangle_A$ and $|0\rangle_B$ in Eq. (3) are the vacuum states of A and B annihilated by $a$ and $b$, respectively: $a|0\rangle_A = 0$, $b|0\rangle_B = 0$. $\theta$ in Eq. (4) is a real parameter. It is possible to introduce a complex parameter, in general, but we take a real one for the later purpose. Note that $U(\theta)$ is a nonlocal operator that entangles A and B.

It is known [6] that two-mode squeezed states are related to thermal states. Let us see this point in terms of an operation of the form in Eq. (1). Ignore the mode of B. Then, the reduced density matrix of the subsystem A is given by the partial trace over B:

$$\rho(A) = \text{Tr}_B\left[|\theta\rangle\langle\theta|\right]. \tag{5}$$

If the partial trace is performed by the use of the number-state basis $\left\{|n\rangle_B = \left(b^\dagger\right)^n |0\rangle_B / \sqrt{n!}\right\}_{n=0,1,2,\ldots}$, then Eq. (5) is written in the form of Eq. (1):

$$\rho(A) = \Lambda(\rho_0(A)) = \sum_{n=0}^{\infty} K_n \rho_0(A) K_n^\dagger, \tag{6}$$

where

$$\rho_0(A) = |0\rangle_{AA}\langle 0|, \tag{7}$$

$$K_n = {}_B\langle n|U(\theta)|0\rangle_B.  \quad (8)$$

To calculate $K_n$, it is convenient to decompose $U(\theta)$ as follows [7]:

$$U(\theta) = \exp(a^\dagger b^\dagger \tanh\theta)\exp\left[-(a^\dagger a + b^\dagger b + 1)\ln(\cosh\theta)\right]\exp(-ab\tanh\theta). \quad (9)$$

With this form, $K_n$ is immediately calculated to be

$$K_n = \frac{(a^\dagger \tanh\theta)^n}{\sqrt{n!}\,\cosh\theta}\exp\left[-a^\dagger a \ln(\cosh\theta)\right]. \quad (10)$$

The trace-preserving condition in Eq. (2) is fulfilled, and $\{K_n^\dagger K_n\}_{n=0,1,2,\ldots}$ forms a POVM. However, the operation is not unital [8], since $\sum_{n=0}^\infty K_n K_n^\dagger = I_A/\cosh^2\theta \neq I_A$, that is, $I_A$ is not a fixed point: $\Lambda(I_A) \neq I_A$.

$\rho(A) = \Lambda(\rho_0(A))$ is expressed as follows:

$$\rho(A) = \sum_{n=0}^\infty p_n |n\rangle_{AA}\langle n|, \quad (11)$$

$$p_n = \frac{(\tanh^2\theta)^n}{\cosh^2\theta}. \quad (12)$$

Eq. (11) shows that perfect decoherence is realized because of the absence of off-diagonal terms. The value of the von Neumann entropy, $S[\rho] = -\text{Tr}(\rho\ln\rho)$, vanishes for the pure state $\rho = \rho_0(A)$ in Eq. (7), whereas it is nonvanishing for $\rho = \rho(A) \equiv \Lambda(\rho_0(A))$ in Eq. (11). However, in the next section, the operation in Eq. (6) is shown to be invertible. In particular, if $\theta$ is chosen to be

$$\cosh\theta = \frac{1}{\sqrt{1-\exp(-\beta\omega_A)}}, \quad (13)$$

then $\rho(A)$ in Eq. (11) becomes the canonical density matrix with the inverse temperature $\beta$:

$$\rho(A) = \frac{1}{Z(\beta)} \exp(-\beta H_A), \quad (14)$$

where $H_A = \omega_A a^\dagger a$ is the Hamiltonian of $A$ with the frequency $\omega_A$ and $Z(\beta)$ the partition function given by $Z(\beta) = \text{Tr}_A \exp(-\beta H_A) = 1/[1-\exp(-\beta\omega_A)]$.

Closing this section, we mention that thermofield dynamics [9] gives a basis for how the subdynamics of the two-mode squeezed state generates the thermal state, and a quantum-optical implication of this fact is discussed in Ref. [6]. However, the choice in Eq. (13) is not necessarily made in the subsequent discussion, and a general form in Eq. (11) is considered.

### III. PHASE-SPACE REPRESENTATION OF COMPLETELY-POSITIVE QUANTUM OPERATION: OPERATION ARISING FROM SUBDYNAMICS OF TWO-MODE SQUEEZED VACUUM STATE AND ITS INVERTIBILITY

Our interest is in the structure of Eq. (6) with Eq. (10). In this section, we discuss its phase-space representation in order to "visualize" such an abstract operation. For this purpose, let us recall the phase-space operator [5]

$$\Delta^{(s)}(\alpha,\alpha^*) = \frac{1}{\pi^2} \int d^2z \, \exp\left[-\frac{s}{2}z^*z + (a^\dagger - \alpha^*)z - z^*(a-\alpha)\right], \quad (15)$$

where $\alpha$ is the complex phase-space variable, $d^2z \equiv d(\text{Re}\,z)d(\text{Im}\,z)$, and the integration is performed over the whole complex $z$-plane. It satisfies the relations

$$\int d^2\alpha \, \Delta^{(s)}(\alpha,\alpha^*) = I_A, \quad (16)$$

$$\text{Tr}_A\left[\Delta^{(s)}(\alpha,\alpha^*)\Delta^{(-s)}(\alpha',\alpha'^*)\right] = \frac{1}{\pi}\delta^2(\alpha-\alpha'). \quad (17)$$

where $\delta^2(\alpha) \equiv \delta(\text{Re}\,\alpha)\delta(\text{Im}\,\alpha)$. $\Delta^{(s)}(\alpha,\alpha^*)$ defines the correspondence relation between a quantum operator and its classical counterpart in phase space:

$$Q_s(a,a^\dagger) = \int d^2\alpha \, Q(\alpha,\alpha^*) \Delta^{(s)}(\alpha,\alpha^*), \quad (18)$$

$$Q(\alpha,\alpha^*) = \pi \, \text{Tr}_A\left[Q_s(a,a^\dagger)\Delta^{(-s)}(\alpha,\alpha^*)\right], \quad (19)$$

where $Q_s(a,a^\dagger)$ is the $s$-ordered operator obtained by quantizing a classical quantity $Q(\alpha,\alpha^*)$. Three special cases are important: $s=0$, $s=-1$, and $s=+1$ correspond to the Weyl ordering, normal ordering, and anti-normal ordering, and the phase-space distribution

$$F^{(s)}(\alpha, \alpha^*) = \text{Tr}_A\left[\rho(A)\, \Delta^{(s)}(\alpha, \alpha^*)\right] \tag{20}$$

becomes the Wigner distribution function, Sudarshan-Glauber $P$ function, and Husimi $Q$ function, respectively. Among them, the Husimi $Q$ function is always positive semidefinite, whereas the others can take negative values, in general. From Eq. (16), $F^{(s)}(\alpha, \alpha^*)$ is seen to be normalized: $\int d^2\alpha\, F^{(s)}(\alpha, \alpha^*) = 1$. Also, from Eqs. (18) and (20), a quantum expectation value is expressed as a statistical average in phase space:

$$\text{Tr}_A\left[\rho(A)\, Q_s(a, a^\dagger)\right] = \int d^2\alpha\, Q(\alpha, \alpha^*)\, F^{(s)}(\alpha, \alpha^*). \tag{21}$$

To formulate the phase-space representation of the quantum operation in Eq. (1), the $P$ representation with $s = -1$ turns out to be convenient. In this representation, the density matrix, $\rho = \rho(A)$, is expressed as follows:

$$\rho(A) = \int d^2\alpha\, P(\alpha, \alpha^*) |\alpha\rangle_{AA}\langle\alpha|, \tag{22}$$

where $P(\alpha, \alpha^*) \equiv F^{(-1)}(\alpha, \alpha^*)$, and $|\alpha\rangle_A \equiv \exp(a^\dagger\alpha - a\alpha^*)|0\rangle_A$ is the coherent state that is the eigenstate of the annihilation operator, $a|\alpha\rangle_A = \alpha|\alpha\rangle_A$, satisfying the

(over-)completeness relation, $(1/\pi)\int d^2\alpha |\alpha\rangle_{AA}\langle\alpha| = I_A$. Then, the $P$ representation of Eq. (1) is found to have the form of an integral transformation:

$$P(\alpha,\alpha^*) = \int d^2\alpha' \, G(\alpha,\alpha^* : \alpha',\alpha'^*) P_0(\alpha',\alpha'^*), \tag{23}$$

where $P$ and $P_0$ are the $P$ representations of $\rho(A)$ and $\rho_0(A)$, respectively. $G(\alpha,\alpha^* : \alpha',\alpha'^*)$ is the transformation kernel given by

$$G(\alpha,\alpha^* : \alpha',\alpha'^*) = \sum_n {}_A\langle\alpha'|K_n^\dagger \Delta^{(-1)}(\alpha,\alpha^*) K_n |\alpha'\rangle_A, \tag{24}$$

which satisfies $\int d^2\alpha \, G(\alpha,\alpha^* : \alpha',\alpha'^*) = 1$, as can be seen from Eqs. (2) and (16). Eq. (23) with Eq. (24) is the formula for the $P$ representation of a general completely-positive quantum operation.

Now, we evaluate the kernel for the operators in Eq. (10). Performing a straightforward calculation, we find it to be given by

$$G(\alpha,\alpha^* : \alpha',\alpha'^*) = \frac{1}{\pi \sinh^2\theta} \exp\left(-\frac{1}{\sinh^2\theta}|\alpha - \alpha'\cosh\theta|^2\right). \tag{25}$$

Thus, in this case, the transformation kernel in the $P$ representation is Gaussian. [In the special case when the choice in Eq. (13) is made, then the factor, $\sinh^2\theta = 1/[\exp(\beta\omega_A)-1]$, which is the internal energy of $A$ in an equilibrium state with inverse temperature $\beta$, is responsible for the thermal broadening of the width of the Gaussian function.] What to be noted is its $\theta$-dependent translational invariance: $\alpha \to \alpha + \xi$, $\alpha' \to \alpha' + \xi/\cosh\theta$, where $\xi$ is an arbitrary complex constant. In the operator form in Eqs. (6) and (10), it would be hard to see such a symmetry. This invariance enables one to rewrite Eq. (23) as a convolution of the kernel in Eq. (25) and an arbitrary initial distribution:

$$P(\alpha, \alpha^*) = \int d^2\alpha' \, G(\alpha - \alpha', \alpha^* - \alpha'^*) \tilde{P}_0(\alpha', \alpha'^*), \qquad (26)$$

where

$$G(\alpha - \alpha', \alpha^* - \alpha'^*) = \frac{1}{\pi \sinh^2\theta} \exp\left(-\frac{1}{\sinh^2\theta}|\alpha - \alpha'|^2\right), \qquad (27)$$

$$\tilde{P}_0(\alpha, \alpha^*) = \frac{1}{\cosh^2\theta} P_0\left(\frac{\alpha}{\cosh\theta}, \frac{\alpha^*}{\cosh\theta}\right). \qquad (28)$$

Therefore, using the standard technique of the Laplace-Fourier transformation, we can conversely express $P_0$ in terms of $P$, implying a remarkable fact that the quantum operation in Eq. (1) with Eq. (10), which maps from the vacuum pure state at vanishing temperature to a mixed state with perfect decoherence, is invertible.

Finally, noting that the $P$ representation of the "initial" vacuum state in Eq. (7) is given by

$$P_0(\alpha, \alpha^*) = \delta^2(\alpha), \qquad (29)$$

it immediately follows from Eq. (26) that

$$P(\alpha, \alpha^*) = \frac{1}{\pi \sinh^2 \theta} \exp\left(-\frac{1}{\sinh^2 \theta}|\alpha|^2\right), \tag{30}$$

which precisely reproduces the *P* representation of the thermal state in Eq. (11).

## IV. CONCLUDING REMARKS

We have developed the phase-space representations of completely-positive quantum operations and described them as integral transformations in phase space in order to describe such abstract operations in terms of a more intuitively perceivable classical notions. We have studied as an example the *P* representation of the operation associated with the nonunitary subdynamics of the two-mode squeezed state, which maps from the vacuum state of the reduced single-mode field at vanishing temperature to the thermal state. We have derived the explicit form of the kernel of the integral transformation in phase space and found that the translational invariance of the kernel in phase space is revealed, leading to a remarkable result that the operation is invertible, and coherence lost by the quantum operation can be recovered. This highlights usefulness of the phase-space representations.

Quantum-optical techniques are widely used in quantum information processing, where errors occur due to decoherence [3]. The present result may have significance for e.g. error correction, since lost coherence in the single-mode state arising from the total two-mode state can be restored because of the invertibility discovered here.

We expect that the phase-space representations can offer a useful method for revealing properties of quantum operations of various kinds in a way similar to the present one. For example, it is of interest to extend the discussion to the case of systems in finite Hilbert space [10].

**ACKNOWLEDGMENT**

S. A. was supported in part by a Grant-in-Aid for Scientific Research from the Japan Society for the Promotion of Science and by the Ministry of Education and Science of the Russian Federation (federal program of competitive growth of Kazan Federal University).

______________________________